\documentclass[prl,nobalancelastpage,twocolumn,superscriptaddress,nolongbibliography]{revtex4-2}
\pdfoutput=1

\usepackage{color,amsthm,amsmath,amsxtra,amsfonts,dsfont,graphicx,bm,amssymb}
\usepackage[colorlinks=true,linkcolor=blue, citecolor=blue, urlcolor=blue, bookmarks]{hyperref}
\usepackage{centernot}
\usepackage[dvipsnames]{xcolor}
\usepackage{graphicx}
\usepackage{tikz}
\usepackage{braket}
\usepackage{natbib}
\usepackage{url}
\usepackage{hyperref}

\usepackage{fancyhdr}

\newcommand{\be}{\begin{equation}}
\newcommand{\ee}{\end{equation}}
\newcommand{\bea}{\begin{eqnarray}}
\newcommand{\eea}{\end{eqnarray}}
\newcommand{\bse}{\begin{subequations}}
\newcommand{\ese}{\end{subequations}}
\newcommand{\opnorm}[1]{\left\|#1\right\|}

\theoremstyle{plain}

\newtheorem{prop}{Proposition}[section]

\newtheorem{lem}{Lemma}[section]

\newcommand{\prlsection}[1]{{\em {#1}.---~}}

\begin{document}
\title{ Hamiltonian $k$-Locality is the Key Resource for Powerful Quantum Battery Charging }

\author{Anupam \surname{Sarkar}}
\email{asarkar@imsc.res.in}
\author{Sibasish \surname{Ghosh}}
\email{sibasish@imsc.res.in}
\affiliation{Optics and Quantum Information Group, The Institute of Mathematical Sciences,
CIT Campus, Taramani, Chennai 600113, India.}
\affiliation{Homi Bhabha National Institute, Training School Complex, Anushakti Nagar, Mumbai 400094, India.}

\begin{abstract}
Storing and extracting energy using quantum degrees of freedom is a promising approach to leveraging quantum effects in energy science. Early experimental efforts have already demonstrated its potential to surpass the charging power of existing technologies. In this context, it is crucial to identify the specific quantum effects that can be exploited to design the most efficient quantum batteries and push their performance to the ultimate limit. While entanglement has often been considered a key factor in enhancing charging (or discharging) power, our findings reveal that it is not as critical as previously thought. Instead, three parameters emerge as the most significant in determining the upper bound of instantaneous charging power: the locality of the battery and charger Hamiltonians, and the maximum energy storable in a single unit cell of the battery. To derive this new bound, we have also addressed several open questions previously noted in the literature but lacks an explanation. This bound provides a foundation for designing the most powerful charger-battery systems, where combined optimization of both components offers enhancements that cannot be achieved by manipulating only one of them.
\end{abstract}

\maketitle


\prlsection{Introduction} The transition from being mere spectators of wildfire to harnessing and using fire in daily life marks a significant milestone in the history of humanity's ability to manipulate energy in a controlled manner~\cite{Carnot1824}. The potential sources of energy we utilize are deeply tied to the length and time scales over which we can exercise precise control. Unsurprisingly, advances in our ability to manipulate systems at the atomic scale~\cite{bluv2021,daley2022} have opened new avenues for leveraging quantum phenomena in energy science. One such promising concept is the quantum battery \cite{Alicki2013}, which has seen remarkable progress in recent years, both theoretically and experimentally \cite{Quach2023,Campai2024}.

A quantum battery is composed of multiple subsystems, or unit cells, where each subsystem is a quantum system, such as a single electron or atom with precisely controlled degrees of freedom, such as spin or energy levels. This approach fundamentally differs from the best available energy storage technologies today \cite{Kim2019}, where energy storage and extraction rely predominantly on ion migration through electrolytes (solid or liquid). Despite numerous challenges \cite{Gherardini2020,Quach2020,Francica2020,Shi2022,crescente2020charging,garcia2020fluctuations,bakhshinezhad2024tradeoffs,Castellano2024} that need to be addressed before quantum batteries can become a practical technology, their potential for delivering high charging (and discharging) power and superior energy density \cite{quach2022,metzler2023emergence} compared to conventional methods warrants significant attention from the scientific community.

Existing battery-charger models in the literature \cite{Binder2015,Campaioli2017} can broadly be classified into two categories: one involving an interacting battery and a non-interacting charger Hamiltonian \cite{Le2018,Dou2022,Zhang2024,Ghosh2020,Zhao2021}, and the other involving a non-interacting battery and an interacting charger\cite{Rossini2020,Ferraro2018,Catalano2024,Julia-Farre2020}. In this letter, we present a unified mathematical framework capable of studying both scenarios. Using this framework, we propose a new bound in proposition \ref{prop2} on the charging power for systems where both the battery and charger are interacting. Remarkably, we observe an enhancement in charging power that cannot be achieved when either the battery or the charger is non-interacting. To uncover the origin of this enhancement, we demonstrate that entanglement within the system or generated during evolution plays no significant role in determining the charging power, an issue that has been the subject of considerable attention in the literature \cite{Andolina2019,Andolina2019b,Julia-Farre2020,Kamin2020,Zhang2023,Gyhm2024}. Instead, three key parameters drive this enhancement: $1.$ The \textit{locality} of the battery Hamiltonian, $2.$ The \textit{locality} of the charger Hamiltonian, and $3.$ The maximum energy that can be stored in a single unit cell.

Our motivation stems from two key observations. First, prior studies \cite{Gyhm2022} have shown that the participation number plays a negligible role in determining the charging power of a quantum battery. Additionally, our numerical investigations \cite{Sarkar2025} reveal that the extensive power advantage predicted in \cite{Campaioli2017} does not manifest in long-range, all-to-all connected models, even when both the charger and battery Hamiltonians are interacting. Second, in long-range interacting systems, a stronger form of locality emerges \cite{Kuwahara2016a,Chen2019,Kuwahara2020,Tran2020}, characterized by Lieb-Robinson-like bounds on information spreading. This lack of a power advantage in long-range, all-to-all connected models further hints at the presence of a stronger notion of locality in the energy dynamics of these systems, which we term locality of energy. Specifically, we demonstrate that even in the presence of entanglement, energy under the action of a local operator remains localized rather than spreading across the entire lattice, facilitated by system correlations. This phenomenon also clarifies observations made in \cite{Le2018}, where charging only the boundary sites of an interacting battery had minimal impact on energy distribution within the system's bulk.

We introduce the $g$-extensive condition \cite{Arad2016}, which ensures fair comparison between any \textit{parallel} and \textit{interacting} charging protocol, particularly in long-range, all-to-all connected models. Physically, this condition enforces a finite energy capacity at each lattice site. Its significance in bounding charging power also elucidates the observation \cite{Ghosh2022} that increasing the Hilbert space dimension of the battery’s subsystems/unit cells enhances charging power. Moreover, we provide a constructive proof that any $k$-local $g$-extensive Hamiltonian can be replaced by an energetically equivalent Hamiltonian comprising commuting terms, achieving the same maximum power. This result demonstrates that unitary circuit-based charging protocols are as powerful as Hamiltonian-based charging protocols, nullifying any power advantage due to non-commutativity of the Hamiltonian terms \cite{Campaioli2017}.

\prlsection{Locality and $g$-extensiveness of Hamiltonian}
We consider a lattice model for both the quantum battery and its charger. At each lattice site, a unit cell of Hilbert space with dimension $d$ is assumed, capable of storing and exchanging energy. The collection of all such unit cells is denoted as $\Lambda={1, 2, \cdots, N}$. The Hamiltonians of the charger and the battery are assumed to be $k$-local and $q$-local respectively, where these parameters determine the maximum number of sites involved in any single interaction term:
\begin{equation}
H^{C}= \sum_{|X|\leq k}h_{X}^{C}\quad\text{and}\quad H^{B}=\sum_{|X|\leq q}h_{X}^{B}
\end{equation}
where $X$ can represent any subset of $\Lambda$. This form accommodates both short-range and long-range interactions, as the lattice sites contained in subsets $X\subset\Lambda$ are not necessarily geometrically local.
An additional condition imposed on both $H^{B}$ and $H^{C}$ is $g$-extensivity:
\begin{equation}
\sum_{X:i\in X}||h_{X}^{B/C}||\leq g\quad \forall i\in\Lambda
\end{equation}
where $g$ is a constant of order $\mathcal{O}(1)$, independent of the lattice size.  The $g$-extensivity condition ensures that the energy associated with each unit cell is bounded by a finite value $g$. In long-range, all-to-all connected models, this condition guarantees the extensive property of energy for both the battery and charger Hamiltonians, providing a natural framework for fair comparisons between different charging protocols, including parallel charging schemes. For a detailed discussion of this issue and how $g$-extensivity addresses it, refer to our accompanying paper \cite{Sarkar2025}. It is clear from this context that this parameter $g$ determines the energy scale of the system and in the following sections we will show how the locality parameters $k$ and $q$ dictate the timescale for energy transformation.

With this framework established for the battery and charger Hamiltonians, we can now analyze the charging dynamics. The battery is initially in a state $\ket{\psi(0)}$ with average energy $\langle E(0)\rangle=\bra{\psi(0)}H^{B}\ket{\psi(0)}$. The goal of charging is to excite the battery to a higher energy state within a finite time interval $T$, achieving an energy $\langle E(T)\rangle=\bra{\psi(T)}H^{B}\ket{\psi(T)}$. Two quantities of primary interest in the literature for quantifying charging power are the \textit{average power}, $\langle P(T)\rangle=\frac{E(T)-E(0)}{T}$, and the \textit{instantaneous power}, $\left|P(t)\right|=|\frac{d}{dt}E(t)|$. We focus on the instantaneous power, as the average power satisfies $\langle P(t)\rangle\leq\max_{t}P(t)$, providing an upper bound on the latter automatically applies to the former. Using the equation of motion \cite{SM}, the instantaneous power can be upper bounded by the quantity, $\left|P(t)\right|\leq||[H^{C},H^{B}]||$. Applying the triangle inequality and the sub-multiplicative property of the operator norm yields:
\begin{equation}\label{eq5}
\left|P(t)\right|\leq||[H^{C},H^{B}]||\leq2||H^{C}||.||H^{B}||
\end{equation}
This bound is considered trivial as it fails to capture the locality of interactions \cite{Hastings2010} inherent to any physical quantum many-body systems. Since energy is an extensive variable, $||H^{B}||$ and $||H^{C}||$ scale with the system size, suggesting that power could be increased indefinitely by enlarging the battery or charger. A similar limitation arises in bounding the minimum time for state transformation in quantum many body system using traditional energy-time uncertainty relations \cite{Bukov2019,Lam2021,Chen2023}.

Analogous to the Lieb-Robinson bound \cite{Lieb1972}, where locality restricts instantaneous information propagation across the lattice, we demonstrate that locality imposes fundamental limits on instantaneous energy transfer in the battery, irrespective of system size or interaction structure. These limits arise from the interplay between the locality parameters ($k$, $q$) and the energy scale ($g$), providing a more physical description of the charging dynamics.

\prlsection{Locality of Energy Dynamics} While the locality of the charger Hamiltonian has already been established \cite{Gyhm2022} as a significant factor in determining the upper bound on charging power, there is no direct conclusion regarding the role of entanglement. In this work, leveraging an important mathematical result \cite{Arad2016}, which we refer to as the Arad-Kuwahara-Landau-Hastings (AKLH) lemma \cite{SM}, we argue that the entanglement does not play a constructive role in determining the upper bound on charging power. This observation aligns with the findings in the case of mixed states \cite{Campaioli2017}. To understand how the locality and $g$-extensive property of the Hamiltonian strongly restrict energy transitions in the battery, and consequently impose stricter bounds on $P(t)$ compared to Equation \ref{eq5}, we consider the quantity $||\Pi_{[m\epsilon,m\epsilon+\epsilon)}h_{X}^{C}\Pi_{[m'\epsilon,m'\epsilon+\epsilon)}||$. Here, $\Pi_{[m\epsilon,m\epsilon+\epsilon)}$ is the projection operator onto the energy eigenspace of the battery corresponding to eigenvalues within the interval $[m\epsilon, m\epsilon+\epsilon)$, where $\epsilon > 0$ and $m \in \mathbb{Z}$. The term $h_{X}^{C}$ represents a local term in the charger Hamiltonian, quantifying the maximum possible energy transition in the battery induced by this local term. This quantity plays a central role in our analysis as it provides an upper bound on instantaneous energy transitions in the battery under the action of the charger Hamiltonian. Furthermore, it helps to clarify the role of the participation number :
\begin{equation}
\resizebox{\columnwidth}{!}{$
\begin{aligned}[t]
||\Pi_{[m\epsilon,m\epsilon+\epsilon)}&H^{C}\Pi_{[m'\epsilon,m'\epsilon+\epsilon)}|| = ||\Pi_{[m\epsilon,m\epsilon+\epsilon)}\sum_{|X|\leq k}h_{X}^{C}~\Pi_{[m'\epsilon,m'\epsilon+\epsilon)}||\\
& \leq \sum_{X}||\Pi_{[m\epsilon,j\epsilon+\epsilon)}h_{X}^{C}\Pi_{[m'\epsilon,m'\epsilon+\epsilon)}||.
\end{aligned}
$}
\end{equation}

\begin{lem}
\label{lem:uniqueLemma}
The maximum energy transition in the battery under any local term present in the charger Hamiltonian is bounded as, 
\[||\Pi_{[m\epsilon,m\epsilon+\epsilon)}h_{X}^{C}\Pi_{[m'\epsilon,m'\epsilon+\epsilon)}|| \leq ||h_{X}^{C}|| e^{-\frac{1}{2gq}[(m-m')\epsilon-\epsilon - 4A]}\]
\end{lem}

To interpret the term $A$ in the exponential factor, consider the case where $h_{X}^{C}$ acts on $k$ lattice sites, $S_{1} = \{i_{1}, i_{2}, \dots, i_{k}\} \subset \Lambda$. The battery Hamiltonian also contains local terms, such as $h_{Y}^{B}$, which have non-trivial support on at least one lattice site in $S_{1}$. Collecting all such terms from the battery Hamiltonian forms the set $\mathcal{E}_{X}^{B}$. Then, $A = \sum\limits_{h_{Y}^{B} \in \mathcal{E}_{X}^{B}} ||h_{Y}^{B}||$. For a detailed proof of this lemma in the context of quantum batteries, refer to \cite{SM}. Here, we aim to illustrate the consequences of this lemma by discussing some examples of battery-charger models considered in the literature \cite{Campai2024}.

Consider the following example to illustrate this concept, Let the battery Hamiltonian be :
\begin{equation}\label{eq8}
H^{B} = h\sum_{i}\sigma_{i}^{z}-\sum_{i<j}g_{ij}[\sigma_{i}^{z}\sigma_{j}^{z}+\alpha(\sigma_{i}^{x}\sigma_{j}^{x}+\sigma_{i}^{y}\sigma_{j}^{y})]
\end{equation}
and the charger Hamiltonian :
\begin{equation}\label{eq9}
H^{C} =\omega\sum_{i=1}^{N}\sigma_{i}^{x}
\end{equation}

The battery Hamiltonian is interacting and 2-local ($q=2$), while the charging Hamiltonian is non-interacting and 1-local ($k=1$). For example, consider the local term $h_{X}^{C} = \omega\sigma_{1}^{x}$, which has non-trivial support on lattice site $1$. Thus, $S_{1} = \{1\}$. Next, identify all local terms in the battery Hamiltonian with non-trivial support on lattice site $1$. These include $h\sigma_{1}^{z}$ and $(N-1)$ terms of the form $-g_{1j}[\sigma_{1}^{z}\sigma_{j}^{z} + \alpha(\sigma_{1}^{x}\sigma_{j}^{x} + \sigma_{1}^{y}\sigma_{j}^{y})]$. The participation number for lattice site $1$ is therefore $N = (N-1) + 1$ for an all-to-all connected model.

Using our notation, the set $\mathcal{E}_{X}^{B}$ contains these $N$ terms, and $A = ||h\sigma_{1}^{z}|| + \sum_{j}||-g_{1j}[\sigma_{1}^{z}\sigma_{j}^{z} + \alpha(\sigma_{1}^{x}\sigma_{j}^{x} + \sigma_{1}^{y}\sigma_{j}^{y})]||$. Due to the $g$-extensive property of the battery, we have $A \leq g$. Substituting into the inequality, we find:
\begin{equation}
\resizebox{\columnwidth}{!}{$
\begin{aligned}[t]
||\Pi_{[m\epsilon,m\epsilon+\epsilon)}\omega\sigma_{1}^{x}\Pi_{[m'\epsilon,m'\epsilon+\epsilon)}|| \leq ||\omega\sigma_{1}^{x}|| e^{-\frac{1}{4g}[(m-m')\epsilon-\epsilon - 4\mathcal{O}(g)]}
\end{aligned}
$}.
\end{equation}

Thus, the maximum possible energy transition induced by this local term is exponentially suppressed beyond the energy range $4g$. For any general $k$-local term in the charger Hamiltonian, the $g$-extensive property implies that the maximum possible energy transition in the battery is of the order $\mathcal{O}(4gk)$ \cite{SM}, beyond which the probability of such transitions is exponentially suppressed.

A simple consequence of this inequality is that increasing $g$, the amount of energy that can be stored in a single lattice site, can significantly affect the charging power. This is because $g$ is a key factor in the exponential term, alongside the locality of the Hamiltonian. The enhancement in power due to the dimensionality of the system is reported in \cite{Ghosh2022}.

This observation also addresses findings in \cite{Le2018}. Considering the same battery and charger Hamiltonian as in \ref{eq8} and \ref{eq9}, it has been observed that connecting the charger at the boundary sites, i.e., $H^{C} = \omega\sigma_{1}^{x} + \omega\sigma_{N}^{x}$, leaves the energy in the bulk unaffected. This can be explained using our framework as follows:
\begin{equation*}
\resizebox{\columnwidth}{!}{$
\begin{aligned}[t]
||\Pi_{[m\epsilon,m\epsilon+\epsilon)}~&H^{C}~\Pi_{[m'\epsilon,m'\epsilon+\epsilon)}||
\leq|\omega|(||\sigma_{1}^{x}||+||\sigma_{N}^{x}||)e^{-\frac{1}{4g}[(m-m')\epsilon-\epsilon-4g]}\\
\end{aligned}
$}
\end{equation*}
Since both local terms in the charging Hamiltonian are $1$-local, this inequality shows that any excitation beyond $4\mathcal{O}(g)$ is exponentially suppressed. Consequently, only nearby boundary sites can be energetically filled, while the bulk sites remain unaffected by the charging process within a short time limit.

The most significant aspect to note here is that even with entanglement present throughout the lattice, the energy transferred from the charger remains localized in the region where the charger is applied. This is particularly striking considering the fact that battery's state is highly entangled, as it suggests that correlations do not facilitate energy propagation across the entire lattice. This indicates that entanglement does not play a significant role in energy dynamics. We term this phenomenon the \textit{locality of energy}.

On the other hand if we consider the battery Hamiltonian to be non-interacting (\(q = 1\)), it can be expressed as  
\[
H^B = h \sum_i \sigma_i^z,
\]
where \(h\) is the local energy term. The charging Hamiltonian is assumed to be all-to-all connected and 2-local, given by  
\[
H^C = \alpha \sum_{i<j} (\sigma_i^x \sigma_j^x + \gamma \sigma_i^y \sigma_j^y) + B \sum_i \sigma_i^z.
\]
For any local term \(h_X^C = \alpha (\sigma_i^x \sigma_j^x)\), lemma~\ref{lem:uniqueLemma} implies that  
\begin{equation*}
\resizebox{\columnwidth}{!}{$
\begin{aligned}[t]
\left\| \Pi_{[m\epsilon, m\epsilon+\epsilon)} ~ \alpha (\sigma_1^x \sigma_2^x) ~ \Pi_{[m'\epsilon, m'\epsilon+\epsilon)} \right\| 
\leq \left\| \alpha (\sigma_1^x \sigma_2^x) \right\| e^{-\frac{1}{2g}[(m-m')\epsilon-\epsilon- 4.2g]},
\end{aligned}
$}
\end{equation*}
where \(A = \|h \sigma_1^z\| + \|h \sigma_2^z\| \leq 2g\). There are a total of \(\frac{N(N-1)}{2} \approx \mathcal{O}(N^2)\) 2-local terms in the charger Hamiltonian. Since \(\|\sigma_1^x \sigma_2^x\| = 1\), any such term in the Hamiltonian contributes to the exponential factor as \(e^{-\frac{1}{2g}[(m-m')\epsilon-\epsilon - 4.2g]}\).  

Using the triangle inequality, we find that  
\begin{equation*}
\resizebox{\columnwidth}{!}{$
\begin{aligned}[t]
\left\| \Pi_{[m\epsilon, m\epsilon+\epsilon)} H_x^C \Pi_{[m'\epsilon, m'\epsilon+\epsilon)} \right\| 
\leq \alpha \mathcal{O}(N^2) e^{-\frac{1}{2g}[(m-m')\epsilon-\epsilon - 4.2g]},
\end{aligned}
$}
\end{equation*}
where \(H_x^C = \alpha \sum_{i<j} \sigma_i^x \sigma_j^x\). A similar bound holds for \(H_y^C = \alpha \sum_{i<j} \sigma_i^y \sigma_j^y\).  

For the $1$-local terms in the Hamiltonian, we have  
\begin{equation*}
\resizebox{\columnwidth}{!}{$
\begin{aligned}[t]
\left\| \Pi_{[m\epsilon, m\epsilon+\epsilon)} B \sum_i \sigma_i^z \Pi_{[m'\epsilon, m'\epsilon+\epsilon)} \right\| 
\leq |B| N e^{-\frac{1}{2g}[(m-m')\epsilon-\epsilon - 4g]}.
\end{aligned}
$}
\end{equation*}
To ensure that the charger's energy is extensive, a normalization factor proportional to \(\frac{1}{N}\) must be incorporated. This normalization ensures that the pre-factor of the exponential term scales as \(\mathcal{O}(N)\), which matches the scaling for parallel charging scenario.  

Alternatively, if the charger Hamiltonian comprises highly non-local terms with non-trivial support on \(\mathcal{O}(N)\) sites, the exponential factor becomes \((m-m')\epsilon-\epsilon \approx 4\mathcal{O}(Ng)\). This configuration enables the fastest possible energy transitions, providing a power advantage scaling as \(\mathcal{O}(N)\) compared to the parallel charging case, where \((m-m')\epsilon-\epsilon \approx 4\mathcal{O}(g)\).  

Thus, we conclude that the charging advantage cannot grow with system size unless non-local operations involving all battery sites are applied, while maintaining the extensivity of the charger's energy. In \cite{Gyhm2022}, a similar observation is reported under the condition on the charger Hamiltonian, where \(H^C = \sum_{j,m} h_{jm} \ket{E_j} \bra{E_m}\) and \(h_{jm} = 0\) if \(|E_j - E_m| > \Delta E\), with \(\{\ket{E_j}\}\) being the energy eigenstates of the battery Hamiltonian.

We provide an alternative proof \cite{SM} of bounding the maximum power by assuming the charger Hamiltonian is \(k\)-local and \(g\)-extensive as these conditions automatically imposes constraints on transition probabilities. Transitions beyond \(\mathcal{O}(4gk)\) for a \(k\)-local Hamiltonian are exponentially suppressed but not exactly zero.  

\begin{prop}\label{prop1}
For a \( g \)-extensive, non-interacting battery, charging through a \( k \)-local, interacting charger \( H^{C} \), which satisfies the extensive condition of energy, the instantaneous charging power is bounded as
\[
|P(t)| \leq 12gk \| H^{C} \|.
\]
\end{prop}
This proposition suggests that even in an all-to-all connected model for the charger Hamiltonian, at any given instance during the charging process, only one of the $k$-local links becomes active, facilitating the energy transfer to the battery. While a specific spin can participate to multiple $k$-local terms, at any moment, only one of these terms is relevant for injecting energy. Consequently, the increasing non-commutativity between local terms in all-to-all connected models, which is related to the participation number \cite{Campaioli2017} has no impact on the instantaneous charging power (see figure $1$). This indicates that a \textit{Unitary circuit-based charging} protocol is equally effective as a \textit{Hamiltonian-based charging} protocol. Indeed, in \cite{SM}, we demonstrated  that any $k$-local, extensive charging Hamiltonian is energetically equivalent to a Hamiltonian composed of commuting terms.
\begin{figure}[htbp]
    \centering
    \includegraphics[width=0.4\textwidth]{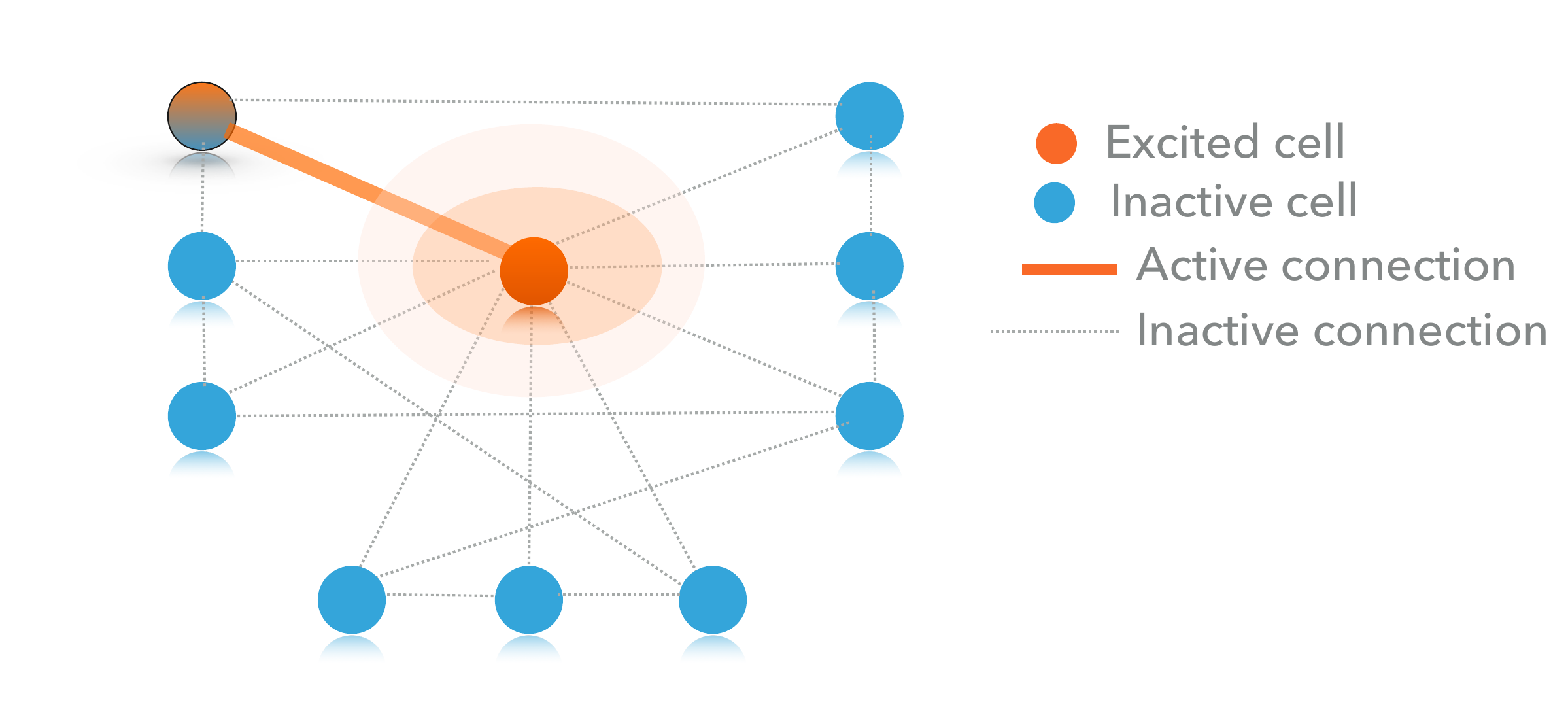}
    \caption{A schematic to represent the charging process: Here the Hamiltonian is $2$-local and the central spin is connected to the charger, although it participates in many interaction as represented by the connected edges. At any instant only one of those links gets activated and helps the corresponding connected spin to be energetically filled. }
    \label{fig:your-label}
\end{figure}

While Proposition \ref{prop1} establishes an upper bound on the charging power for a non-interacting battery under an interacting charger protocol, it completely neglects the influence of the locality of the battery Hamiltonian. On the other hand, a boost in charging power has been observed \cite{Le2018,Dou2022,Zhang2024} when considering an interacting battery. To bridge this gap and provide a bound that incorporates both factors, we demonstrate \cite{SM} that the locality of the battery Hamiltonian can be just as critical as the locality of the charger Hamiltonian. This observation will be formally stated as follows:
\begin{prop}\label{prop2}
    If the battery and charger Hamiltonian both are interacting and respectively \(q\)-local and \(k\)-local and both satisfies the $g$-extensive condition of energy then the instantaneous charging power is bounded as,
    \[
    |P(t)|\leq 12gkq||H^{C}||
    \]
\end{prop}

\prlsection{Conclusion and Discussion}Our work clarifies the role of entanglement and Hamiltonian locality in enhancing the charging power of quantum batteries. Specifically, we demonstrate that the entanglement generated during the charging process is a by-product of the interaction terms in the Hamiltonian, not the other way around. Specifically, in the case of an interacting battery, we showed that even if the initial or intermediate state during charging is entangled, a non-interacting charger fails to utilize this entanglement to spread energy efficiently. Instead, the energy remains localized in regions where the charger is directly applied. This observation is consistent with prior studies \cite{Keating2015,Kim2013} showing that in spin-chain systems, entanglement and energy exhibit distinct behaviors during system evolution.

A key finding of our work is the identification of three primary factors that determine the maximum charging power: (1) the $g$-extensivity of energy (a bound on the energy of a single unit cell), (2) the locality of the battery Hamiltonian \((q)\), and (3) the locality of the charger Hamiltonian \((k)\). Together, these factors provide a framework for understanding and optimizing charging power. Our results suggest that bosonic systems, which can theoretically hold unbounded energy at a single site, are ideal candidates for designing high-power quantum batteries. This claim is supported by recent experimental evidence \cite{quach2022} as well as theoretical studies \cite{Andolina2024} highlighting the potential of bosonic batteries and chargers.

Regarding the question of which states are most advantageous for maximizing charging power, we conjecture that the locally irreversible states \cite{Kuwahara2017} are optimal. This is because their energy variance scales super-extensively with system size as well as these states contain macrscopic superposition. Investigating the relationship between Hamiltonian locality, energy variance and it's effect in charging power \cite{Banuls2020,Kuwahara2016,Anshu2016, Rai2024} would be an exciting direction for future research.

Drawing an analogy with the quantum search algorithm, where entanglement is not necessary for achieving a speedup over classical methods \cite{Lloyd1999}, we raise a key question: is it possible to identify a pathway in the quantum state space that enhances charging power beyond the non-interacting case without generating any entanglement? Such a pathway could offer deeper insights, akin to established results in contexts such as initial mixed states \cite{Campaioli2017} and work extraction \cite{Hovhannisyan2013}.

\prlsection{Acknowledgments}
A.S thanks Tomotaka Kuwahara for kindly granting permission to name the lemma in this work as per our choice.

%


\onecolumngrid
\appendix
\section*{Supplemental Material}  
\section{Bounding Charging power using operator norm}\label{sec:charging}


The average energy of the battery at time $t$ is given by \[\langle E_{B}\rangle_{t}=\bra{\psi(t)}H^{B}\ket{\psi(t)}\]. 
The time evolution of the state under the charging Hamiltonian $H^C$ is given by $\ket{\psi(t)}=e^{-iH^{C}t}\ket{\psi(0)}$ (taking $\hbar=1$). Therefore, $\langle E_{B}\rangle_{t}=\bra{\psi(0)}e^{+iH^{C}t}H^{B}e^{-iH^{C}t}\ket{\psi(0)}$ The instantaneous rate of change of the average energy, which represents the charging power, is given by 
\[ \begin{aligned} 
P(t)=\frac{d}{dt}\langle E_{B}\rangle_{t}&=\bra{\psi(0)}iH^{C}e^{iH^{C}t}H^{B}e^{-iH^{C}t}-ie^{iH^{C}t}H^{B}e^{-iH^{C}t}H^{C}\ket{\psi(0)}\\ 
&=i\bra{\psi(0)}e^{iH^{C}t}[H^{C},H^{B}]e^{-iH^{C}t}\ket{\psi(0)}\\ 
&=i\bra{\psi(t)}[H^{C},H^{B}]\ket{\psi(t)} \end{aligned}\] 

To bound this instantaneous power, we utilize the operator norm. For any operator $A$, the operator norm is defined as: \[||A||_{\infty}:=\max\limits_{\substack{\ket{\phi}\in\mathcal{H}\\||\phi||=1}}||A\ket{\phi}||=\max\limits_{\substack{\ket{\phi},\ket{\psi}\in\mathcal{H}\\||\phi||=||\psi||=1}}\left|\bra{\psi}A\ket{\phi}\right|\]

Where \(||A\ket{\phi}||^{2}:=\bra{\phi}A^{\dagger}A\ket{\phi}\) defines the norm of the state $A\ket{\phi}$. Using this definition of operator norm and the expression for $P(t)$ derived above, we can bound the magnitude of instantaneous power: \[\left|P(t)\right|=\left|\frac{d}{dt}\langle E_{B}\rangle_{t}\right|=\left|\bra{\psi(t)}i[H^{C},H^{B}]\ket{\psi(t)}\right|\leq\left|\left|[H^{C},H^{B}]\right|\right|\]

This bound is quite general and extends naturally to mixed states $\rho(t)$. In that case, the power can be derived from the quantum Liouville equation $\frac{d\rho}{dt}=-i[H^C,\rho]$, leading to the same upper bound: \[\left|P(t)\right|=\left|\text{Tr}(H^B\frac{d\rho}{dt})\right|=\left|\text{Tr}(H^B[-i[H^C,\rho]])\right|=\left|\text{Tr}(\rho[H^C,H^B])\right|\leq\left|\left|[H^C,H^B]\right|\right|\]

It is easy to see now that the average charging power is bounded by maximum instantaneous power as follows,
\[\langle P(T)\rangle=\frac{1}{T}\int_{0}^{T}\frac{d}{dt}\langle E_{B}\rangle_{t}dt\leq\frac{1}{T}\int_{0}^{T}\left|\left|[H^{C},H^{B}]\right|\right|dt=\left|\left|[H^{C},H^{B}]\right|\right|\]

\section{ Lattice model of Battery and Charger Hamiltonian}
We consider a \(D\)-dimensional lattice \(\Lambda\subset\mathbb{Z}^{D}\) containing \(N\) sites indexed as \(\Lambda={1,2,\cdots, N}\). At each lattice site, we place a \(d\)-dimensional quantum system that serves as an elementary unit capable of storing and exchanging energy. We refer to each such quantum system as a {\it{unit cell}} forming the fundamental building blocks of quantum battery

The battery and charger systems are described by \(q\)-local and \(k\)-local Hamiltonians respectively, chosen to capture the essential physics while maintaining mathematical tractability. These Hamiltonians take the form:

\[H^{B}=\sum_{|X|\leq q}h_{X}^{B}\quad H^{C}=\sum_{|X|\leq k}h^{C}_{X}\]

where \(X\) represents any subset of \(\Lambda\), and \(h_X^{B/C}\) are operators acting on the subset \(X\). The parameters \(k\) and \(q\) are constants independent of the lattice size, and \(|X|\) denotes the cardinality of subset \(X\). This general formulation allows us to represent both short-range and long-range interactions within a unified framework.

\subsection{Example} For short-range interactions, such as nearest-neighbor Ising coupling, a typical local term takes the form \(\sigma_{i}^{z}\sigma_{i+1}^{z}\), where \(\sigma_{i}^{z}\) represents the Pauli \(Z\) operator acting on lattice site \(i\) for a spin \(\frac{1}{2}\) system. In this case, the subsets \(X\) are pairs of adjacent sites: \(\{1,2\},\{2,3\},\cdots,\{i,i+1\},\cdots\).

For long-range interactions, \(X\) can be any arbitrary two-element subset, such as \(\{1,2\},\{1,7\},\{1,i\}, \{2,8\}\\,
\{2,j\},\cdots\), allowing for interactions between spatially far separated sites.

We consider time-independent Hamiltonians for both battery and charger systems, as these prove optimal for charging under bounded energy bandwidth conditions \cite{}. This physical constraint reflects that unbounded energy would allow instantaneous state transitions, which is unrealistic in practical implementations.

\section{\(g\)-extensive condition}Both battery and charger Hamiltonians must satisfy the \(g\)-extensive condition \cite{Arad2016}:

\[\sum_{X:i\in X}\left|\left|h_{X}^{B/C}\right|\right|\leq g\quad i\in\Lambda\]

where \(g\) is a finite constant representing the maximum energy that can be stored in or transferred through a single unit cell. This condition ensures proper scaling of energy with system size and prevents unphysical energy accumulation in individual cells.

In non-interacting parallel charging schemes \cite{Campaioli2017}, the total available energy in the charger is bounded by \(gN\), reflecting the maximum energy \(g\) per unit cell. Therefore, \(||H^{C}||\leq gN\).

For all-to-all connected charger Hamiltonians (such as SYK model or LMG model), the number of local terms \(h^{C}_{X}\) scales as \(\mathcal{O}(N^2)\). To maintain extensive energy scaling comparable to the parallel charging case, the \(g\)-extensive condition provides natural normalization:

\[||H^{C}||=\left|\left|\sum_{X}h^{C}_{X}\right|\right|\leq\sum_{X}\left|\left|h^{C}_{X}\right|\right|=\sum_{i=1}^{N}\sum_{X:~i\in X}\left|\left|h^{C}_{X}\right|\right|\leq\sum_{i=1}^{N}g=gN\]

This ensures that long-range interactions do not provide an unfair energetic advantage compared to parallel charging schemes while maintaining the desired extensive scaling of the total energy with system size.

\section{ Arad-Kuwahara-Landau-Hastings (AKLH) lemma and locality of energy} 
Consider an arbitrary state of the battery\(\ket{\psi}\) expressed as a superposition of eigenstates of \(H^B\) with energies in the interval \([0,E]\). The {\it{AKLH lemma}} provides bounds on how an arbitrary operator affects this state. Specifically, let \(\Pi_{[0,E]}^B\) be the projection operator onto the energy eigenspace of the battery Hamiltonian. Then for the initial battery state \(\ket{\psi}\), we have  \(\ket{\psi}=\Pi_{[0,E]}^{B}\ket{\psi}\), where \(E > 0\) can be chosen arbitrarily. In practice, \(E\) can be taken to be small since the battery typically starts in a low-energy state before charging.

A fundamental question arises: What is the effect of an arbitrary local term \(h^C_X\) of the charger Hamiltonian on this state? More specifically, how much can this local term change the state's energy? The AKLH lemma addresses these questions through the following precise statement:
\begin{lem}
For any \(g\)-extensive, \(q\)-local battery Hamiltonian \(H^{B}\), if \(\Pi_{[0,E]}^{B}\) and \(\Pi_{[E',\infty]}^{B}\) are the projection operators onto the eigenspace of \(H^{B}\)whose energies are \(<E\) and \(>E'\) respectively then,

\[\left|\left|\Pi_{[E',\infty]}^{B}h^{C}_{X}\Pi_{[0,E]}^{B}\right|\right|\leq||h^{C}_{X}||\exp\left[-\frac{1}{2gq}(\Delta E-4A)\right]\]

where \(\Delta E=E'-E\) and for the operator \(h_{X}^{C}\), let \(\mathcal{E}_{X}^{B}\) be a subset of interaction terms in the battery Hamiltonian such that, 

\[[H^{B},h^{C}_{X}]=\sum_{Y\in \mathcal{E}_{X}^{B}}[h^{B}_{Y},h_{X}^{C}]\quad \text{and}\quad A=\sum_{Y\in \mathcal{E}_{X}^{B}}\left|\left|h_{Y}^{B}\right|\right|\]

 \(\mathcal{E}_{X}^{B}\) represents the terms in the battery Hamiltonian which are non-commuting with a specific local term \(h^{C}_{X}\) in the  charger Hamiltonian. 
\end{lem}
For a \(k\)-local term \(h^C_X\) with non-trivial support on lattice sites \(\{i_{1},i_{2},\cdots i_{k}\}\) , this lemma reveals that energy transitions under such a local operator are exponentially suppressed beyond \(4gK\). This result has profound implications: even when \(\ket{\psi}\) is highly entangled, the correlations present in the state do not enable energy to spread across the entire lattice, a behavior that parallels classical systems. Consequently, the short-time excitations in the battery are predominantly determined by the non-trivial support of the charger term rather than the correlation properties of the state.

This locality of energy explains why long-range and short-range models exhibit similar energetic behavior, offering no significant advantages in the latter case, despite their markedly different entanglement properties.\\

While this lemma was originally established in \cite{Arad2016}, we present here a detailed proof using notation consistent with our framework, specifically adapted for quantum battery applications.

\vspace{2mm}
{\bf{Proof:}} 
\[
\left|\left|\Pi_{[E',\infty]}^{B}h_{X}^{C}\Pi_{[0,E]}^{B}\right|\right|=\left|\left|\Pi_{[E',\infty]}^{B}e^{-pH^{B}}e^{pH^{B}}h_{X}^{C}e^{-pH^{B}}e^{pH^{B}}\Pi_{[0,E]}^{B}\right|\right|
\] 
Using Cauchy-Scwarz inequality and from the definition of operator norm we can write,
\[
\begin{aligned}
\left|\left|\Pi_{[E',\infty]}^{B}h_{X}^{C}\Pi_{[0,E]}^{B}\right|\right|&\leq\left|\left|\Pi_{[E',\infty]}^{B}e^{-pH^{B}}\right|\right|~\left|\left|e^{pH^{B}}h_{X}^{C}e^{-pH^{B}}\right|\right|~\left|\left|e^{pH^{B}}\Pi_{[0,E]}^{B}\right|\right|\\
&\leq e^{-p(E'-E)}\left|\left|e^{pH^{B}}h_{X}^{C}e^{-pH^{B}}\right|\right|
\end{aligned}
\]
Now to upper bound the quantity \(\left|\left|e^{pH_{B}}h_{X}^{C}e^{-pH_{B}}\right|\right|\), we first examine how the locality of the operator \(e^{pH^{B}}h_{X}^{C}e^{-pH^{B}}\) grows. We employ the Hadamard formula:

\[
e^{pH^{B}}h_{X}^{C}e^{-pH^{B}}=\sum\limits_{n=0}^{\infty}\frac{p^{n}}{n!}\mathcal{C}_{n}
\]

where \(\mathcal{C}_{n}\) denotes the \(n\)-fold nested commutator defined as:

\[
\mathcal{C}_{n}=\underbrace{[H^{B},[H^{B},[\cdots[H^{B},h_{X}^{C}]\cdots]]]}_{\text{$n$ times}}
\]

Here, \(p\) is a free parameter.

To analyze the locality of a generic nested commutator term, we first examine \(\mathcal{C}_{1}=[H^{B},h_{X}^{C}]\). Given that the battery Hamiltonian is \(q\)-local and \(h_{X}^{C}\) has non-trivial support on \(k\) lattice sites, \([H^{B},h_{X}^{C}]\) is \((k+q)\)-local. The commutator \([H^{B},h_{X}^{C}]\) receives contributions only from terms in \(H^{B}\) that have non-trivial support on at least one of these \(k\) sites; all other terms vanish. We denote the set of such contributing terms from \(H^{B}\) as \(\mathcal{E}_{X}^{B}\), and define \(A=\sum\limits_{Y\in\mathcal{E}_{X}^{B}}\|h_{Y}^{B}\|\).

Then the second commutator \(\left[H^{B},[H^{B},h_{X}^{C}]\right]\) includes terms which have non-overlapping support with at least one of the terms in \(Y\in \mathcal{E}_{X}^{B}\) or \(h_{X}^{C}\). Let's denote this set as \(\mathcal{E}_{X_{1}}^{B}\). The maximum locality of any term in this commutator will be \((k+2q)\). For instance, for a \(2\)-local battery Hamiltonian, a generic term in the nested commutator can be \(\left[h_{12},[h_{14},h'_{1}]\right]\) and the locality of this term is bounded by \((1+2\times2)=5\). Since the second commutator introduces an operator which has maximum non-trivial support on \(q\) new sites and due to the \(g\)-extensive condition, the operator norm will be bounded by \((A+gq)\). The operator norm of a generic \(n\)-th commutator term is upper bounded by the quantity \([A+(n-1)gq]\), that is:

\[\sum\limits_{Y\in\mathcal{E}^{B}_{X_{n-1}}}||h^{B}_{Y}||\leq\left[A+(n-1)gq\right]\]
Now,
\[
\begin{aligned}
\left|\left|[H_{B},h^{C}_{X}]\right|\right|=\left|\left|\left[\sum\limits_{Y\in\mathcal{E}_{X}^{B}}h_{Y}^{B},h_{X}^{C}\right]\right|\right|\leq\sum\limits_{Y\in\mathcal{E}_{X}^{B}}\left|\left|[h_{Y}^{B},h_{X}^{C}]\right|\right|\leq 2~||h_{X}^{C}||\sum\limits_{Y\in\mathcal{E}_{X}^{B}}||h_{Y}^{B}||
\end{aligned}
\]
It is easy to see from the argument above 
\[
||~\mathcal{C}_{2}~||=\left|\left|\left[H_{B},\left[H_{B},h^{C}_{X}\right]\right]\right|\right|\leq2^{2}~||h_{X}^{C}||\sum\limits_{Y_{1}\in\mathcal{E}_{X_{1}}^{B}}\sum\limits_{Y\in\mathcal{E}_{X}^{B}}||h_{Y_{1}}^{B}||.||h_{Y}^{B}||
\]
Therefore for the $n$-th nested commutator we can write, 
\[
\begin{aligned}
||~\mathcal{C}_{n}~||&\leq 2^{n}~||h_{x}^{C}||\sum\limits_{Y_{n-1}\in\mathcal{E}_{X_{n-1}}^{B}}\sum\limits_{Y_{n-2}\in\mathcal{E}_{X_{n-2}}^{B}}\cdots\sum\limits_{Y\in\mathcal{E}_{X}^{B}}||h_{Y_{n-1}}^{B}||.||h_{Y_{n-2}}^{B}||\cdots||h_{Y}^{B}||\\
&=2^{n}~||h_{x}^{C}||\sum\limits_{Y_{n-1}\in\mathcal{E}_{X_{n-1}}^{B}}||h_{Y_{n-1}}^{B}||\sum\limits_{Y_{n-2}\in\mathcal{E}_{X_{n-2}}^{B}}||h_{Y_{n-2}}^{B}||\cdots\sum\limits_{{Y}\in\mathcal{E}_{X}^{B}}||h_{Y}^{B}||\\
&=2^{n}~||h_{x}^{C}||~\left[A+(n-1)gq\right].\left[A+(n-2)gq\right]\cdots\left[A+gq\right].A\\
&=(2gq)^{n}~||h_{X}^{C}||\left[\frac{A}{gq}.(1+\frac{A}{gq})\cdots\left\{(n-1)+\frac{A}{gq}\right\}\right]
\end{aligned}
\]
To get a non-local term whose supports cover the whole lattice we need $\lfloor\frac{N}{q}\rfloor$ nested commutators. 

Therefore,

\[
\begin{aligned}
\Big|\Big|e^{pH_{B}}h_{X}^{C}e^{-pH_{B}}\Big|\Big|&=\Big|\Big|\sum_{n=0}^{\infty}\frac{p^{n}}{n!}\mathcal{C}_{n}\Big|\Big|\leq\sum_{n=0}^{\infty}\frac{p^{n}}{n!}||~\mathcal{C}_{n}~||\\
&\leq\sum_{n=0}^{\infty}\frac{p^{n}}{n!}(2gq)^{n}\Big[\frac{A}{gq}.(1+\frac{A}{gq})\cdots\{(n-1)+\frac{A}{gq}\}\big].||h_{X}^{C}||\\
\textrm{We define}~\frac{A}{gq}=a,\\
&=\sum_{n=0}^{\infty}\frac{(2pgq)^{n}}{n!}\Big[\big((n-1)+a\big).\big((n-2)+a\big).\cdots(1+a).a\Big].||h_{X}^{C}||\\
&=\frac{||h_{X}^{C}||}{(1-2pgq)^{a}}
\end{aligned}
\]
Where we have used the series expansion of the quantity $\frac{1}{(1-2pgq)^{a}}$ under the condition $0<2pgq<1$.

Therefore 
\[
\left|\left|\Pi_{[E',\infty]}^{B}h_{X}^{C}\Pi_{[0,E]}^{B}\right|\right|\leq e^{-p(E'-E)}\frac{||h_{X}^{C}||}{(1-2pgq)^{a}}
\]
We have to minimize this quantity with respect to the free parameter $p$ so 

\[
\frac{d}{dp}\left[\frac{e^{-p(E'-E)}}{(1-2pgq)^{a}}\right]=0\] gives us ,
\[
p=\frac{1}{2gq}\left[1-\frac{2agq}{\Delta E}\right]=\frac{1}{2gq}\left[1-\frac{2A}{\Delta E}\right]
\]

Therefore \(e^{-p\Delta E}=e^{-\frac{1}{2gq}\left[{\Delta E}-2A\right]}\)

On the other hand  $(1-2pgq)^{-a}=\left[\frac{2A}{\Delta E}\right]^{-\frac{A}{gq}}$, by writing this term in exponential form we get,

\[\left|\left|\Pi_{[E',\infty]}^{B}h_{X}^{C}\Pi_{[0,E]}^{B}\right|\right|\leq ||h_{X}^{C}||.e^{-\frac{1}{2gq}\left[\Delta E-2A-2A\ln{\frac{\Delta E}{2A}}\right]}\]

Using the condition $\Delta E\geq 0$ and the inequality $\ln(x)\geq1-\frac{1}{x}$ in fact we write it in more compact form, \[\left|\left|\Pi_{[E',\infty]}^{B}h_{X}^{C}\Pi_{[0,E]}^{B}\right|\right|\leq ||h_{X}^{C}||.e^{-\frac{1}{2gq}\left[\Delta E-4A\right]}
\]

\section{Bounding the power using AKLH lemma}
\section{Non-interacting Battery and Interacting Charger}
The eigenvalues of the battery Hamiltonian are real and we shall assume that it is bounded in the interval $[E_{max},E_{min}]$. The eiegenvalues are discrete points in this interval for finite dimensional systems. To apply the lemma directly in our context we discretize the whole interval by unit $\epsilon$  by using the following decomposition,

\[H^{B'}=\sum_{m=-\infty}^{\infty}(m+\frac{1}{2})\epsilon\,\Pi_{[m\epsilon,m\epsilon+\epsilon)}\]

Where $\Pi_{[m\epsilon,m\epsilon+\epsilon)}$ is the projector onto the eigenspace of $H_{B}$ of energy eigenvalues lies in the interval $[m\epsilon,m\epsilon+\epsilon)$. If any of the eigenvalue of $H_{B}$ lies in the interval $[j\epsilon,j\epsilon+\epsilon)$ then the projector is non-zero, otherwise it is zero. We can take $\epsilon$ to be as small as possible. According to this decomposition, the corresponding eigenvalue is the mean of this interval i.e $\frac{m\epsilon+(m+1)\epsilon}{2}=(m+\frac{1}{2})\epsilon$.

On account of this discretization process we have introduced an error in the eigenvalue as it can be the case that the eigenvalue of $H^{B}$ actually coincides with the limit points $m\epsilon$ or $(m+1)\epsilon$ of the interval but in our discretized Hamiltonian we have considered the eigenvalue $(m+\frac{1}{2})\epsilon$, which is in the middle of the interval. Therefore we have introduced an error of amount maximum $\frac{\epsilon}{2}$ in this process. 

To take account of this error we introduce an error Hamiltonian $\delta H_{B}$ and we can write,

\[H^{B}=H^{B'}+\delta H^{B}\] 

It is clear from the above discussion that

\[||\delta H^{B}||\leq\frac{\epsilon}{2}\]

\begin{equation}\label{eq1}
\left|\left|[H^{C},H^{B}]\right|\right|=\left|\left|[H^{C},H^{B'}+\delta H^{B}]\right|\right|\leq\left|\left|[H^{C},H^{B'}]\right|\right|+\left|\left|[H^{C},\delta H^{B}]\right|\right|
\end{equation}
The second term of the equation can be bounded as,
\[\left\|[H^{C},\delta H^{B}]\right\|\leq 2\left|\left|H^{C}\right|\right|\,\left|\left|\delta H^{B}\right|\right|\leq\epsilon\left|\left|H^{C}\right|\right|\]
Now for any state \(\ket{\psi}\), we can write in a superposition of energy eigenbasis of the Hamiltonian $H^{B'}$ as,
\[
\ket{\psi}=\sum_{m\in\mathbb{Z}}\Pi_{[m\epsilon,m\epsilon+\epsilon)}^{B'}\ket{\psi}\]
Therefore we can write,
\[
\begin{aligned}
\opnorm{[H^{C},H^{B'}]}&=\max_{\ket{\psi}}\left|\bra{\psi}[H^{C},H^{B'}]\ket{\psi}\right|\\
&=\max_{\ket{\psi}}\left|\sum_{m,m'}\bra{\psi}\Pi_{[m\epsilon,m\epsilon+\epsilon)}^{B'}(H^{C}H^{B'}-H^{B'}H^{C})\Pi_{[m'\epsilon,m'\epsilon+\epsilon)}^{B'}\ket{\psi}\right|
\end{aligned}
\]
Since $\Pi_{[m\epsilon,m\epsilon+\epsilon)}^{B'}$ is projector on the eigenbasis of $H^{B'}$ we can write,
\[
\begin{aligned}
&=\max_{\ket{\psi}}\left|\sum_{m,m'}(m'-m)\bra{\psi}\Pi_{[m\epsilon,m\epsilon+\epsilon)}^{B'}H^{C}\Pi_{[m'\epsilon,m'\epsilon+\epsilon)}^{B'}\ket{\psi}\right|\\
&=\max_{\ket{\psi}}\left|\sum_{m,m'}(m'-m)\bra{\psi}\Pi_{[m\epsilon,m\epsilon+\epsilon)}^{B'}\Pi_{[m\epsilon,m\epsilon+\epsilon)}^{B'}H^{C}\Pi_{[m'\epsilon,m'\epsilon+\epsilon)}^{B'}\Pi_{[m'\epsilon,m'\epsilon+\epsilon)}^{B'}\ket{\psi}\right|\\
\text{we have used the fact}~\Pi_{[m'\epsilon,m'\epsilon+\epsilon)}^{B'}~&\text{is projector}\\ 
&\leq\sum_{m,m'}|m-m'|\epsilon\opnorm{\bra{\psi}\Pi_{[m\epsilon,m\epsilon+\epsilon)}^{B'}}\opnorm{\Pi_{[m\epsilon,m\epsilon+\epsilon)}^{B'}H^{C}\Pi_{[m'\epsilon,m'\epsilon+\epsilon)}^{B'}}\opnorm{\Pi_{[m'\epsilon,m'\epsilon+\epsilon)}^{B'}\ket{\psi}}
\end{aligned}
\]
Note the term \(\opnorm{\bra{\psi}\Pi_{[m\epsilon,m\epsilon+\epsilon)}^{B'}}\) is the probability amplitude of $\ket{\psi}$ in the energy interval $[m\epsilon,m\epsilon+\epsilon)$, we call it $\alpha_{m}$ and $\alpha_{m'}$ for the last term in the inequality. Since we have assumed $\ket{\psi}$ is normalized therefore $\sum\limits_{m}\alpha_{m}^{2}=1$. 

Therefore \[
\begin{aligned}
\opnorm{[H^{C},H^{B'}]}&\leq\sum\limits_{m,m'}|m-m'|\epsilon~\alpha_{m}\alpha_{m'}\opnorm{\Pi_{[m\epsilon,m\epsilon+\epsilon)}^{B'}H^{C}\Pi_{[m'\epsilon,m'\epsilon+\epsilon)}^{B'}}\\
&\leq\sum\limits_{m,m'}|m-m'|\epsilon~\alpha_{m}\alpha_{m'}H^{C}_{m,m'}
\end{aligned}
\]
Where \(\opnorm{\Pi_{[m\epsilon,m\epsilon+\epsilon)}^{B'}\sum_{X}h^{C}_{X}\Pi_{[m'\epsilon,m'\epsilon+\epsilon)}^{B'}}=H^{C}_{m,m'}\).

This inequality holds significant importance as it governs the probability of the maximum possible energy transition in the battery, driven by the charger Hamiltonian and weighted by the corresponding energy. The locality of the Charger Hamiltonian is crucial, as it serves as the primary factor in bounding such energy transitions from above. Applying {\it{AKLH Lemma}} directly to this quantity yields, 
\begin{equation}\label{eq9}
\opnorm{[H^{C},H^{B'}]}\leq\sum_{X}\sum\limits_{m,m'}|m-m'|\epsilon~\alpha_{m}\alpha_{m'}\opnorm{h_{X}^{C}}e^{-\frac{1}{2gq}\left[(m\epsilon-m'\epsilon+\epsilon)-4A\right]}
\end{equation}
The locality of the charger Hamiltonian is manifested in the quantity $A$ in the exponential factor and correspondingly bounding the maximum allowable energy transition. To see that, consider a battery Hamiltonian is of the form 
\[
H^{B}=h\sum_{i}\sigma_{i}^{z}
\]
and a charger hmailtonian of the form \(H^{C}=\alpha\sum\limits_{i<j}(\sigma_{i}^{x}\sigma_{j}^{x}+\gamma\sigma_{i}^{y}\sigma_{j}^{y})+B\sum\limits_{i}\sigma_{i}^{z}\). 

This model of charger Hamiltonian was analyzed in \cite{Julia-Farre2020} to demonstrate that entanglement serves as a crucial resource for driving energy transitions and significantly enhancing the charging power. The primary motivation for considering a long-range, all-to-all connected model stems from the participation number, which quantifies the number of interaction terms involving a single lattice site. This participation number increases with lattice size and is regarded in \cite{Campaioli2017, Rossini2020} as a key parameter for boosting charging power. Furthermore, charging through such a Hamiltonian can generate entanglement during the charging process, a feature that distinguishes it from evolution under short-range interacting models.

\vspace{2mm}
To determine the quantity $A$ for such a model, Let's consider a local term in the charger Hamiltonian, $h^{C}_{X}=\alpha(\sigma_{1}^{x}\sigma_{2}^{x})$, which has non-trvial support on two lattice sites $1$ and $2$. Therefore according to the definition, $A=\opnorm{h\sigma_{1}^{z}}+\opnorm{h\sigma_{2}^{z}}$. Now due to $g$-extensiveness of the battery Hamiltonian these two lattice sites can hold maximum $2g=2h$ amount of energy. Therefore in this case $A=2h$. Now there are many such terms in the charger Hamiltonian which also has non-trivial support on lattice sites $1$ and $2$. For example, all the terms of the form $\alpha(\sigma_{1}^{x}\sigma_{j}^{x}),~j>1$ and $\alpha(\sigma_{2}^{x}\sigma_{j}^{x}), ~j>2$ also have non-trivial support on those two lattice sites. In fact there are infact $N$ and $(N-1)$ such two local terms in the charger Hamiltonian which has non-trivial support on lattice site $1$ and $2$ respectively. But all of those terms contribute similarly in the exponential factor, that is $A=2g$.The similar restriction holds true for $\alpha\gamma\sigma_{1}^{y}\sigma_{j}^{y}$. Therefore if we collect all those two local terms $h_{X}^{C}$ in the charger Hamiltonian which has non-trivial support on lattice site $1$ and we demand extensive property of energy for charger Hamiltonian too, which is a physical restriction and to ensure fair comparison with parallel charging. 
Then in \ref{eq9} collecting all those terms we get,
\[
\opnorm{h^{C}_{X_{1}}}\leq g; \quad X_{1}~\text{denotes all those terms which has non trivial support on lattice site}~1
\]
From the above argument it must be clear that for any term in the charger Hamiltonian which has non-trivial support on $k$ lattice sites, then $\max{A}=gk$ for such a term and can drive the energy transition in the battery up to $4gk$, beyond that energy difference, any state transition is exponentially suppressed. Therefore we can write 
\[
\begin{aligned}
&H^{C}_{m,m'}=0 \quad \text{if}~ m\epsilon-(m'\epsilon+\epsilon)>4gk\implies|m-m'|>\frac{4gk}{\epsilon}+1\\
&H^{C}_{m,m'}\leq\opnorm{H^{C}}\quad \text{if} ~ m\epsilon-(m'\epsilon+\epsilon)<4gk\implies|m-m'|>\frac{4gk}{\epsilon}+1
\end{aligned}
 \]
 The last inequality follows from the fact \(\opnorm{\Pi_{[m\epsilon,m\epsilon+\epsilon)}^{B'}H^{C}\Pi_{[m'\epsilon,m'\epsilon+\epsilon)}^{B'}}\leq\opnorm{\Pi_{[m\epsilon,m\epsilon+\epsilon)}^{B'}}\opnorm{H^{C}}\opnorm{\Pi_{[m'\epsilon,m'\epsilon+\epsilon)}^{B'}}\leq\opnorm{H^{C}}\)
 
So we can effectively write the equation \ref{eq9} as,
\[
\begin{aligned}
\opnorm{[H^{C},H^{B'}]}&\leq\epsilon\opnorm{H^{C}}\sum\limits_{|m-m'|\leq\frac{4gk}{\epsilon}+1}|m-m'|~\alpha_{m}\alpha_{m'}\\
&\leq\epsilon\opnorm{H^{C}}\sum\limits_{|m-m'|\leq\frac{4gk}{\epsilon}+1}|m-m'|\frac{\alpha_{m}^{2}+\alpha_{m'}^{2}}{2}\\
\text{as}~0\leq\alpha_{m}^{2},\alpha_{m'}^{2}\leq1\\
&\leq\epsilon\opnorm{H^{C}}\sum\limits_{|m-m'|\leq\frac{4gk}{\epsilon}+1}|m-m'|{\alpha_{m'}^{2}}\\
&\leq\epsilon\opnorm{H^{C}}\sum\limits_{m=-\lfloor\frac{4gk}{\epsilon}+1\rfloor}^{\lfloor\frac{4gk}{\epsilon}+1\rfloor}|m|\sum_{m'}{\alpha_{m'}^{2}}\\
\text{using}\sum\limits_{j=-L}^{L}|j|=L(L+1),~j\in\mathbb{Z}~\text{and}~\sum_{m'}\alpha_{m'}^{2}=1\\
&\leq\epsilon\opnorm{H^{C}}\left\lfloor\frac{4gk}{\epsilon}\right\rfloor\left[\left\lfloor\frac{4gk}{\epsilon}+1\right\rfloor\right]
\end{aligned}
\]
Therefore, \[
\begin{aligned}
\opnorm{\left[H^{C},H^{B}\right]}&=\opnorm{[H^{C},H^{B'}]}+\opnorm{[H^{C},\delta H^{B}]}\leq\epsilon\opnorm{H^{C}}\left\lfloor\frac{4gk}{\epsilon}\right\rfloor\left[\left\lfloor\frac{4gk}{\epsilon}+1\right\rfloor\right]+\epsilon\opnorm{H^{C}}\\
\text{If we choose}~ \epsilon=4gk+\delta\epsilon,~\delta\epsilon>0, \text{then}\\ 
&\leq(12gk+3\delta\epsilon)\opnorm{H^{C}}
\end{aligned}
\]
Letting $\delta\epsilon\to 0$ we get,
\[\left|P(t)\right|\leq12gk\opnorm{H^{C}}\]

\section{Interacting battery and Interacting charger}
\section{Unitary Circuit based charging is as powerful as charging through any Hamiltonian} The previously established upper bound on instantaneous power indicates that the participation number does not influence the bound on charging power significantly. Consequently, it is possible to design the most efficient charger without considering models that feature increasing non-commuting terms with system size, such as in all-to-all connected systems. The physical intuition behind this result is as follows: if the charger Hamiltonian contains local terms  that act on a set of sites , then, even if these lattice sites participate in multiple interaction terms within the Hamiltonian, a single term with sufficient magnitude is enough to saturate the energy capacity of the  unit cells it addresses. This idea can be formalized through the following mathematical framework.

If we consider a general $k$-local Hamiltonian where $k=\mathcal{O}(1)$,  does not vary with the system size. The total  Hamiltonian terms present are of the order $\mathcal{O}(N^{k})$ if we consider the most general all-to-all connected models.

Since the locality of the Hamiltonian terms is the most dominant factor in the energy transition as well as it also determine how many terms of specific degree of locality $k$ are present in the Hamiltonian. We divide the Hamiltonian terms according to their locality, forming a subset of interaction terms of different locality, for example $\{\sigma_{i}^{\alpha}\}$, $\{\sigma_{i}^{\alpha}\otimes\sigma_{j}^{\alpha}\},\cdots\{\sigma_{i_{1}}^{\alpha}\otimes\sigma_{i_{2}}^{\alpha}\otimes\cdots\otimes\sigma_{i_{k}}^{\alpha}\}$. Now these terms contribute differently to the energy transition. 

Now using the concept of `energy units' \cite{Kuwahara2016a} $\bar{h}_{X}^{C}$ we will construct a commuting Hamiltonian which is energetically equivalent to our original Hamiltonian. 
\[\bar{h}_{X}^{C}=\epsilon\frac{h_{X}^{C}}{||h_{X}^{C}||}\]

For instance, consider the charger Hamiltonian is of the following form, 
$$H^{C}=B\sum_{i}\sigma_{i}^{z}-\sum_{i<j}g_{ij}[\sigma_{i}^{z}\sigma_{j}^{z}+\alpha(\sigma_{i}^{x}\sigma_{j}^{x}+\sigma_{i}^{y}\sigma_{j}^{y})]$$

Let's focus on a single lattice site $1$ Accumulate those terms of the Hamiltonian which acts on lattice site $1$ non-trivially. There are $1$-local term like $B\sigma_{1}^{z}$ and $N-1$ two local terms of the form $-g_{1j}[\sigma_{1}^{z}\sigma_{j}^{z}+\alpha(\sigma_{1}^{x}\sigma_{j}^{x}+\sigma_{1}^{y}\sigma_{j}^{y})]$. The restriction of $g$-extensivity on the charger Hamiltonian implies that $\sum\limits_{X:1\in X}||h_{X}^{C}||\leq g$

Now we decompose the original charger Hamiltonian in energy units of $\epsilon$ in the following way, All those terms which contains site $1$ can store maximum of energy let's say $S_{1}$, that is $\sum\limits_{X:1\in X}||h_{X}^{C}||=S_{1}\leq g$. Then we assign $\lceil\frac{S_{1}}{\epsilon}\rceil$ energy units to the lattice site $1$ in our newly constructed Hamiltonian. Similarly we can do for lattice site $2$ also, where the term $-g_{12}[\sigma_{1}^{z}\sigma_{2}^{z}+\alpha(\sigma_{1}^{x}\sigma_{2}^{x}+\sigma_{1}^{y}\sigma_{2}^{y})]$ has been already included in our new Hamiltonian and for lattice site $2$ the terms are of the form $-g_{2j}[\sigma_{2}^{z}\sigma_{j}^{z}+\alpha(\sigma_{2}^{x}\sigma_{j}^{x}+\sigma_{2}^{y}\sigma_{j}^{y})];j\neq1$ for $2$ local terms and $B\sigma_{2}^{z}$ for $1$ local term. Similarly we can define $\sum\limits_{X:2\in X}||h_{X}^{B}||=S_{2}\leq g$ and we assign $\lfloor\frac{S_{2}}{\epsilon}\rfloor$ units of energy for lattice site $2$. The similar construction we can apply for other lattice sites too. Now we just need a representative terms in our new Hamiltonian from these $k$-local subsets. 

For example $\sigma_{1}^{\alpha}\sigma_{2}^{\alpha},\ \alpha=x,y,z$ is enough to account for all the other terms in the Hamiltonian containing $1$, we include $\lfloor\frac{S_{1}}{\epsilon}\rfloor$ number of such terms and for lattice sites $2$, we put $\lfloor\frac{S_{2}}{\epsilon}\rfloor$ number of terms for local term $\sigma_{2}^{z}$. So our newly constructed Hamiltonian would be of the form,
\[
\begin{aligned}
\bar{H}^{C} =\left\lfloor\frac{S_{1}}{\epsilon}\right\rfloor(\sigma_{1}^{\alpha}\sigma_{2}^{\alpha})+\left\lfloor\frac{S_{2}}{\epsilon}\right\rfloor\sigma_{2}^{\beta}+\left\lfloor\frac{S_{3}}{\epsilon}\right\rfloor(\sigma_{3}^{\alpha}\sigma_{4}^{\alpha}) +\left\lfloor\frac{S_{4}}{\epsilon}\right\rfloor\sigma_{4}^{\beta} + \cdots + \left\lfloor\frac{S_{N-1}}{\epsilon}\right\rfloor(\sigma_{N-1}^{\alpha}\sigma_{N}^{\alpha})+\left\lfloor\frac{S_{N}}{\epsilon}\right\rfloor\sigma_{N}^{\beta}
\end{aligned}
\]

The only non-commuting part which still remains is in the term $(\sigma_{1}^{\alpha}\sigma_{2}^{\alpha})$. Now it has three parts, $-g_{12}(\sigma_{1}^{z}\sigma_{2}^{z})$, $-g_{12}\alpha(\sigma_{1}^{x}\sigma_{2}^{x}$) and $-g_{12}\alpha(\sigma_{1}^{y}\sigma_{2}^{y})$. Note that they contribute to the operator norm differently so we put different units of energy according to the their strength. So our new constructed Hamiltonian in terms of $x$, $y$ and $z$ component would be, 

\[
\begin{aligned}
\bar{H}^{C}&=\left\lfloor\frac{S_{1}}{3\epsilon}\right\rfloor(\sigma_{1}^{z}\sigma_{2}^{z})+\left\lfloor\frac{|\alpha|S_{1}}{3\epsilon}\right\rfloor(\sigma_{1}^{x}\sigma_{2}^{x})+\left\lfloor\frac{|\alpha|S_{1}}{3\epsilon}\right\rfloor(\sigma_{1}^{y}\sigma_{2}^{y})+\left\lfloor\frac{S_{2}}{\epsilon}\right\rfloor\sigma_{2}^{z}+\left\lfloor\frac{S_{3}}{3\epsilon}\right\rfloor(\sigma_{3}^{z}\sigma_{4}^{z})+\left\lfloor\frac{|\alpha|S_{3}}{3\epsilon}\right\rfloor(\sigma_{3}^{x}\sigma_{4}^{x})+\left\lfloor\frac{|\alpha|S_{3}}{3\epsilon}\right\rfloor(\sigma_{3}^{y}\sigma_{4}^{y})\\
&+\left\lfloor\frac{S_{4}}{\epsilon}\right\rfloor\sigma_{4}^{z}+\cdots+\left\lfloor\frac{S_{N-1}}{3\epsilon}\right\rfloor(\sigma_{N-1}^{z}\sigma_{N}^{z})+\left\lfloor\frac{|\alpha|S_{N-1}}{3\epsilon}\right\rfloor(\sigma_{N-1}^{x}\sigma_{N}^{x})+\left\lfloor\frac{|\alpha|S_{N-1}}{3\epsilon}\right\rfloor(\sigma_{N-1}^{y}\sigma_{N}^{y})+\left\lfloor\frac{S_{N}}{\epsilon}\right\rfloor\sigma_{N}^{z}
\end{aligned}
\]

Now we can write it as, 
\[
\begin{aligned}
\bar{H}^{C} = & \underbrace{\left[\left\lfloor\frac{S_{1}}{3\epsilon}\right\rfloor(\sigma_{1}^{z}\sigma_{2}^{z})+\left\lfloor\frac{S_{3}}{3\epsilon}\right\rfloor(\sigma_{3}^{z}\sigma_{4}^{z})+\cdots+\left\lfloor\frac{S_{N-1}}{3\epsilon}\right\rfloor(\sigma_{N-1}^{z}\sigma_{N}^{z})\right]}_{\bar{H}_{1}^{C}}\\
&+\underbrace{\left[\left\lfloor\frac{|\alpha|S_{1}}{3\epsilon}\right\rfloor(\sigma_{1}^{x}\sigma_{2}^{x})+\left\lfloor\frac{|\alpha|S_{3}}{3\epsilon}\right\rfloor(\sigma_{3}^{x}\sigma_{4}^{x})+\cdots+\left\lfloor\frac{|\alpha|S_{N-1}}{3\epsilon}\right\rfloor(\sigma_{N-1}^{x}\sigma_{N}^{x})\right]}_{\bar{H}_{2}^{C}}\\
&+\underbrace{\left[\left\lfloor\frac{|\alpha|S_{1}}{3\epsilon}\right\rfloor(\sigma_{1}^{y}\!\otimes\!\sigma_{2}^{y})+\left\lfloor\frac{|\alpha|S_{3}}{3\epsilon}\right\rfloor(\sigma_{3}^{y}\!\otimes\!\sigma_{4}^{y})+\cdots+\left\lfloor\frac{|\alpha|S_{N-1}}{3\epsilon}\right\rfloor(\sigma_{N-1}^{y}\sigma_{N}^{y})\right]}_{\bar{H}_{3}^{C}}+\underbrace{\left[\left\lfloor\frac{S_{2}}{\epsilon}\right\rfloor\sigma_{2}^{z}+\left\lfloor\frac{S_{4}}{\epsilon}\right\rfloor\sigma_{4}^{z}+\cdots+\left\lfloor\frac{S_{N}}{\epsilon}\right\rfloor\sigma_{N}^{z}\right]}_{\bar{H}_{4}^{C}}
\end{aligned}
\]

Note that we were able to write down a $2$-local, $g$-extensive Hamiltonian in a form which is sum of $2$-local $2g$ extensive commuting Hamiltonian. Writing it compactly, 
\[\bar{H}_{C}=\bar{H}_{1}^{C}+\bar{H}_{2}^{C}+\bar{H}_{3}^{C}+\bar{H}_{4}^{C}\]
Using the same construction, for any \(k\)-local, \(g\)-extensive Hamiltonian, we can construct a commuting Hamiltonian consisting of \(k\) terms that is energetically equivalent to the original Hamiltonian and provides the most powerful charging. It has been shown \cite{Kuwahara2016a} that 
\[
\bar{k} = k\left\lfloor \frac{g}{\epsilon} \right\rfloor
\]
commuting Hamiltonians are sufficient to reconstruct the original Hamiltonian. Specifically,
\[
\bar{H}^{C} = \frac{1}{\bar{k}} \sum_{p=1}^{\bar{k}} \bar{H}_{p}^{C}
\]
Where \(\bar{H}_{p}^{C}\) are \(k\)-local and \(gk\)-extensive. By this construction, we introduce an error of 
\[
\opnorm{H^{C} - \bar{H}^{C}} = \mathcal{O}(\epsilon N).
\]
As \(\epsilon\) is chosen to be sufficiently small, this error can be made arbitrarily close to zero.

This construction resolves the open question of why circuit-based charging using non-overlapping gates provides the same power as any general non-interacting Hamiltonian.

The same construction can be applied for an interacting battery where any any $q$-local $g$-extensive battery Hamiltonian can be written as sum of commuting Hamiltonian which is $q$-local and $gq$-extensive ,
\[
H^{B}=\frac{1}{\bar{q}}\sum\limits_{p=1}^{\bar{q}}\bar{H}_{p}^{B}
\]
Therefore, \[
\begin{aligned}
\opnorm{[H^{C},H^{B}]}=&\opnorm{\left[H^{C},\frac{1}{\bar{q}}\sum\limits_{p=1}^{\bar{q}}\bar{H}_{p}^{B}\right]}\leq\frac{1}{\bar{q}}\sum\limits_{p=1}^{\bar{q}}\opnorm{[H^{C},\bar{H}_{p}^{B}]}\leq12gkq\opnorm{H^{C}}
\end{aligned}
\]

\end{document}